\documentclass[runningheads]{llncs}

\usepackage[T1]{fontenc}
\def\doi#1{\href{https://doi.org/\detokenize{#1}}{\url{https://doi.org/\detokenize{#1}}}}
\usepackage{graphicx,amsmath,amssymb,bm,subfigure,booktabs,multirow}
\usepackage[misc]{ifsym} 
\usepackage{listings}
\begin{document}
\title{TransEM: Residual Swin-Transformer based regularized PET image reconstruction}
\author{Rui Hu\inst{1}\and 
Huafeng Liu\inst{1,2,3}$^{(\textrm{\Letter})}$} 

\institute{State Key Laboratory of Modern Optical Instrumentation, Department of Optical Engineering,Zhejiang University, Hangzhou 310027, China
\\
\email{liuhf@zju.edu.cn} \and Jiaxing Key Laboratory of Photonic Sensing \& Intelligent Imaging, Jiaxing 314000, China 
\and Intelligent Optics \& Photonics Research Center, Jiaxing Research Institute, Zhejiang University, Jiaxing 314000, China}

\titlerunning{TransEM}
\authorrunning{R. Hu and H. Liu}
\maketitle

\begin{abstract}
Positron emission tomography (PET) image reconstruction is an ill-posed inverse problem and suffers from high level of noise due to limited counts received. Recently deep neural networks especially convolutional neural networks (CNN) have been successfully applied to PET image reconstruction. However, the local characteristics of the convolution operator potentially limit the image quality obtained by current CNN-based PET image reconstruction methods. In this paper, we propose a residual swin-transformer based regularizer (RSTR) to incorporate regularization into the iterative reconstruction framework. Specifically, a convolution layer is firstly adopted to extract shallow features, then the deep feature extraction is accomplished by the swin-transformer layer. At last, both deep and shallow features are fused with a residual operation and another convolution layer. Validations on the realistic 3D brain simulated low-count data show that our proposed method outperforms the state-of-the-art methods in both qualitative and quantitative measures.
\end{abstract}
\section{INTRODUCTION}
Positron Emission Tomography (PET) is one of the irreplaceable tools of functional imaging, which is wildly used in oncology, cardiology, neurology and medical research~\cite{1}. However, PET images usually suffer from high level of noise due to many physical degradation factors and the ill-conditioning of PET reconstruction problem.

To reconstruct high-quality PET images, lots of works have been proposed over the last few decades, which can be roughly divided into five categories: 1)traditional analytic methods such as filtered back-projection (FBP~\cite{2}) and iterative methods like maximum-likelihood expectation maximization (ML-EM~\cite{3}); 2)prior-incorporative methods; 3)image post-processing (denoising) methods; 4) Penalized Log-Likelihood (PLL) methods and 5)deep learning based methods.
\begin{figure}                
\centering
\vspace*{-0pt}
\includegraphics[width=\textwidth]{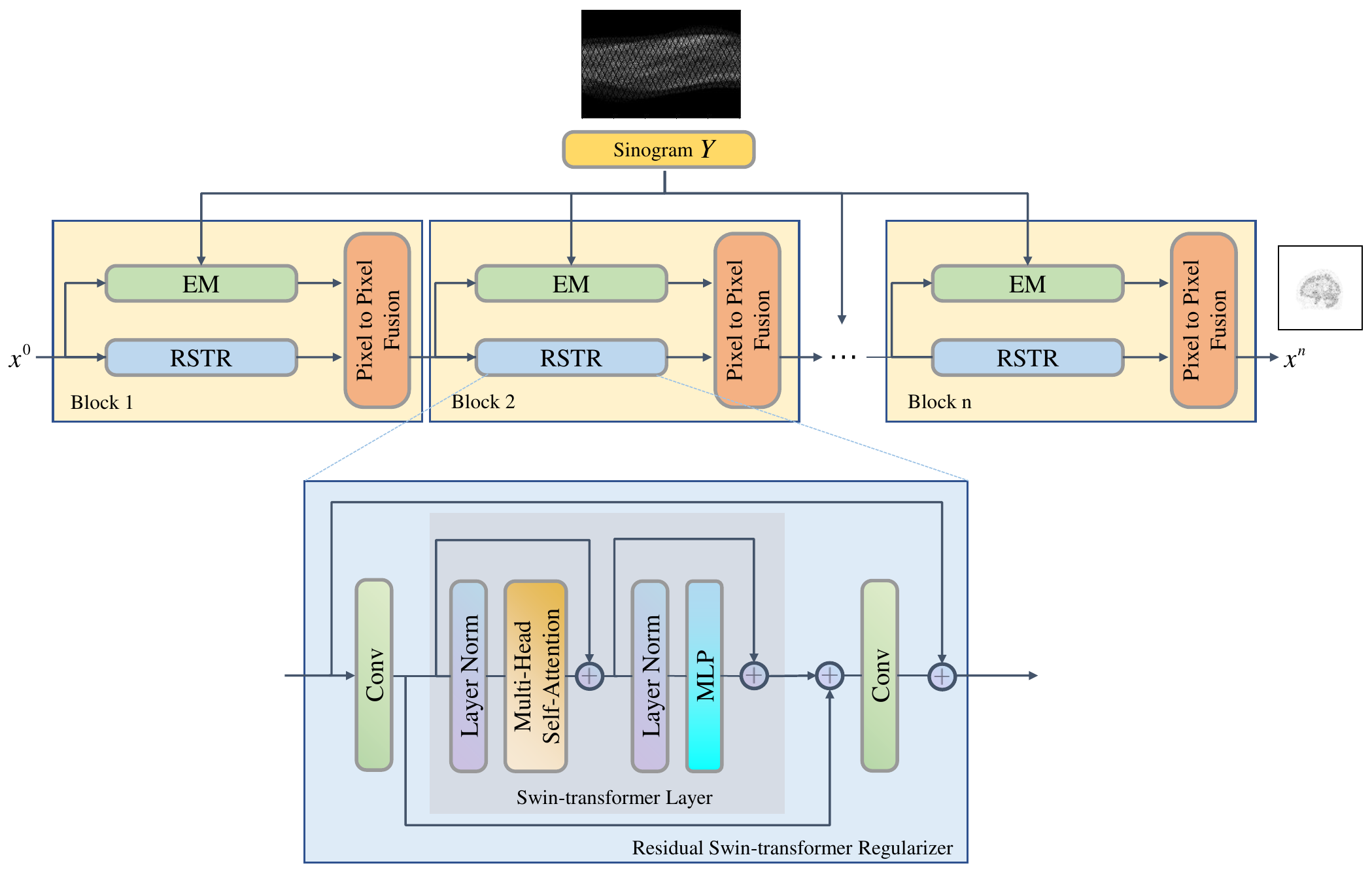}  
\caption{The overall flow-chart of proposed method. Specifically, TransEM is composed of n blocks. Each block contains EM for image updating, RSTR for regularization and a pixel to pixel fusion operation.}  
\label{Fig1}                     
\end{figure}

The FBP algorithm is based on the central slice theorem, which can rapidly finish the reconstruction but suffers from heavy noise due to the lack of modeling of physical properties. Iterative algorithms, such as ML-EM modeled the physical properties and improved image quality. However, the excessive noise propagation from the measurements is the biggest disadvantage of ML solution. To further improve the image quality, prior-incorporative reconstruction methods, image post-processing methods and PLL methods have been introduced. The performance of PLL methods~\cite{PLL,PLL2,PLL3} and prior-incorporative methods like kernel methods~\cite{kernel_method} are closely related to the hyper-parameters that are often hand-crafted before reconstruction. Post-processing is an effective way to reduce noise such as BM3D~\cite{4}, non-local mean (NLM)~\cite{5} and gaussian filter. However, these methods usually tend to be over-smoothing and time-consuming.

Deep learning (DL) techniques especially supervised learning techniques have recently drawn much attention and shown promising results in PET image reconstruction~\cite{7}. Among them, direct learning, DL-based post-denoising and model-based learning are three mainstream approaches. Direct learning~\cite{fbpnet} methods usually learn the mapping from sinogram to the PET image through deep neural networks (DNN). Because there are no physical constraints, direct learning methods are extremely data-hungry and sometimes unstable. DL-based post-denoising methods~\cite{denoise} are simple to implement, but the final results are very sensitive to the pre-reconstruction algorithms. 

By unrolling an iterative reconstruction algorithm, model-based learning shows inspiring results and good interpretability, which has been a promising direction. Gong \emph{et al.} proposed an unrolled network based on 3D U-net and alternating direction method of multipliers (ADMM)~\cite{9}. Mehranian \emph{et al.} proposed a forward backward splitting algorithm for Poisson likelihood and unrolled the algorithm into a recurrent neural network with several blocks~\cite{10}. Lim \emph{et al.} unrolled the block coordinate descent (BCD) algorithm with U-net~\cite{11}. All these methods adopt convolutional neural networks (CNN) to assist in reconstruction. However, a convolution operator has a local receptive field~\cite{12}, giving rise to that CNNs cannot process long-range dependencies unless passing through a large number of layers. while when layer number increases, the feature resolution and fine details may be lost, which limits the quality of reconstructed images. For this issue, the Transformer~\cite{13} is noticed for its strong ability in modeling long-range dependencies of the data and tremendous success in the language domain. Recently, it has also demonstrated promising results in computer vision.

In this paper, we propose a residual swin-transformer~\cite{14} based regularizer (RSTR) along with the ML-EM iterative framework, called TransEM, to reconstruct the standard-dose image from low count sinogram. As one of the model-based learning methods (MoDL), TransEM does not need a large training dataset and achieves state-of-the-art results in realistic 3D brain simulation data.

\section{METHODS AND MATERIALS}
\subsection{Problem formulation}
In PET image reconstruction from sinogram data, The measured data $\bm{y}$ can be well modeled by a Poisson noise model given by:
\begin{equation}
\bm{y} \sim Poisson\{ \overline{\bm{y}}\} \quad s.t. \ \overline{\bm{y}} = \bm{A}\bm{x} + \bm{b} \label{1}
\end{equation}
where $\overline{\bm{y}} \in {{\mathbb R}^I}$ is the mean of the measured data $\bm{y} \in {{\mathbb R}^I}$ with $y_i$ representing the $i$-th detector bin, $\bm{x} \in {{\mathbb R}^J}$ is the unknown activity distribution image with $x_j$ representing $j$-th voxel. $\bm{b} \in {{\mathbb R}^I}$ denotes the expectation of scatters and randoms. $I$ is the number of detector pairs and $J$ is the number of pixels. $\bm{A} \in {{\mathbb R}^{I \times J}}$ is system response matrix with $A_{ij}$ representing the probabilities of detecting an emission from voxel $j$ at detector $i$.

Like many other under-determined inverse problem, the unknown image $\bm{x}$ can be estimated from a Bayesian perspective:
\begin{equation}\label{eq2}
\widehat{\bm{x}} = \mathop {\arg\max}\bm{\limits_x}\ {L(
\bm{y}|\bm{x})} - \beta R(\bm{x})
\end{equation}
\begin{equation}
L(\bm{y}|\bm{x}) = \sum\limits_i {{y_i}\log {{\overline y }_i}}  - {\overline y _i}
\end{equation}
where $L(\bm{y}|\bm{x})$ is the Poisson log-likelihood function of measured sinogram data, $R(\bm{x})$ is the regularization term, $\beta$ is the parameter that controls the regularization.

The forward-backward splitting(FBS) algorithm~\cite{15} and optimization transfer method can be used to solve Eq. (\ref{eq2}). FBS algorithm is used to split the objection function into two terms:
\begin{equation}\label{eq4}
\bm{r}^k = {\bm{x}^{k - 1}} - {\alpha}{\beta}{{\nabla}R}({\bm{x}^{k - 1}})
\end{equation}
\begin{equation}\label{eq5}
{\bm{x}^k} = \mathop {\arg \max }\bm{\limits_x} L(\bm{x}|\bm{y}) - \frac{1}{{2\alpha }}||\bm{x} - {\bm{r}^k}|{|^2}
\end{equation}
where Eq. (\ref{eq4}) is a gradient descent update with step size of $\alpha$ and $k$ denotes $k$-th iteration.
In original FBSEM~\cite{10}, the Eq. (\ref{eq4}) was replaced by a Residual CNN~\cite{16} unit, while the performance of CNN-based regularizer in long-range dependencies is limited due to their localized receptive fields, which limits the quality of the images obtained. To address this issue, we proposed a residual swin-transformer based regularizer (RSTR) to replace the gradient descent update in Eq. (\ref{eq4}):
\begin{equation}
{\bm{r}^k} = RSTR({\bm{x}^{k - 1}})
\end{equation}Eq. (\ref{eq5}) can be reformulated with optimize transfer~\cite{17} method and EM surrogate~\cite{18}:
\begin{equation}\label{eq6}
{\bm{x}^k} = \mathop {\arg \max }\limits_x \sum\limits_j {\hat x_{j,EM}^k\ln ({x_j}) - {x_j} - \frac{1}{{2\alpha \sum\nolimits_i {{A_{ij}}} }}{{(x_j^k - r_j^k)}^2}}
\end{equation}
and $\widehat x_{j,EM}^k$ is given by ML-EM~\cite{3} algorithm:
\begin{equation}\label{eq7}
\hat x_{j,EM}^k = x_j^{k - 1}\frac{1}{{\sum\nolimits_i {{A_{ij}}} }}\sum\nolimits_i {{A_{ij}}} \frac{{{y_i}}}{{{{\overline {{y}}_i }}}}
\end{equation}
setting the derivative of Eq. (\ref{eq6}) to zero, the following closed-form solution can be obtained:
\begin{equation}
x_j^k = \frac{{2x_{j,EM}^k}}{{1 - \frac{{r_j^k}}{{\alpha \sum\nolimits_i {{A_{ij}}} }} + \sqrt {{{(1 - \frac{{r_j^k}}{{\alpha \sum\nolimits_i {{A_{ij}}} }})}^2} + 4\frac{{x_{j,EM}^k}}{{\alpha \sum\nolimits_i {{A_{ij}}} }}} }}
\end{equation}
it can be viewed as a pixel to pixel fusion between regularized reference image $r_{^j}^k$ and ML-EM result $x_{j,EM}^k$. The parameter $\alpha$ was learned from training data.

The whole reconstruction workflow called TransEM is shown in Fig. \ref{Fig1}. The TransEM was unrolled to n blocks, where each block consists of two separate steps and a pixel to pixel fusion operation. The two separate steps are a EM step for image update from measured sinogram data and a deep learning step for prior learning using proposed residual swin transformer based regularizer (RSTR) in image domain.

\subsection{Residual swin-transformer regularizer}
As shown in Fig. \ref{Fig1}, the RSTR is a residual block with a Swin Transformer Layer (STL)~\cite{14} and two convolutional layers. At first, a $3 \times 3$ convolutional layer is used to extract the shallow feature, then a STL is used to extract deep features. At last, another $3 \times 3$ convolutional layer is used to aggregate the shallow and deep features with a residual learning operation.
STL is based on original Transformer layer and multi-head self-attention (MSA), the input of size $H \times W \times C$ is firstly reshaped to a feature map with size of $\frac{{HW}}{{{M^2}}} \times {M^2} \times C$ according to the shifted window mechanism. Then the standard self-attention separately for each window is calculated. After that, a multi-layer perceptron (MLP) with GELU~\cite{Gelu}activation are used. Besides, the residual connection is applied for both modules and the LayerNorm (LN) is added before MLP and MSA.

The whole process of RSTR is formulated as:
\begin{equation}
\begin{gathered}
  {X_1} = Con{v_{3 \times 3}}(Input) \hfill \\
  {X_2} = MSA(LN({X_1})) + {X_1} \hfill \\
  {X_3} = MLP(LN({X_2})) + {X_2} \hfill \\
  Output = Con{v_{3 \times 3}}(X3) + {X_0} \hfill \\ 
\end{gathered}
\end{equation}

\subsection{Implementation details and Reference Methods}
The TransEM was unrolled with ordered subsets (OS) acceleration and implemented using Pytorch 1.7 on a NVIDIA RTX 3090. The number of unrolled Blocks is 60 (10 iterations and 6 subsets). The windowsize of STL (M) is 4. Adam~\cite{19} optimizer and Mean square error (MSE) loss between the network outputs and the label images were used during training. The image $x^0$ was initialized with values of one. The proposed TransEM was compared with conventional ordered subsets expectation maximization (OSEM~\cite{20}), maximum a posterior probability expectation maximization algorithm (MAPEM~\cite{21}), DeepPET~\cite{22} and FBSEM~\cite{10}. For both OSEM and MAPEM, 10 iterations and 6 subsets were adopted. The quadratic penalty was used for MAPEM and the $\beta$ was set to 0.005. Both DeepPET and FBSEM were trained with MSE loss and Adam optimizer. The learning rate was 5e-5, batch size was 4.

\begin{figure}[t]            
\centering
\includegraphics[width=\textwidth]{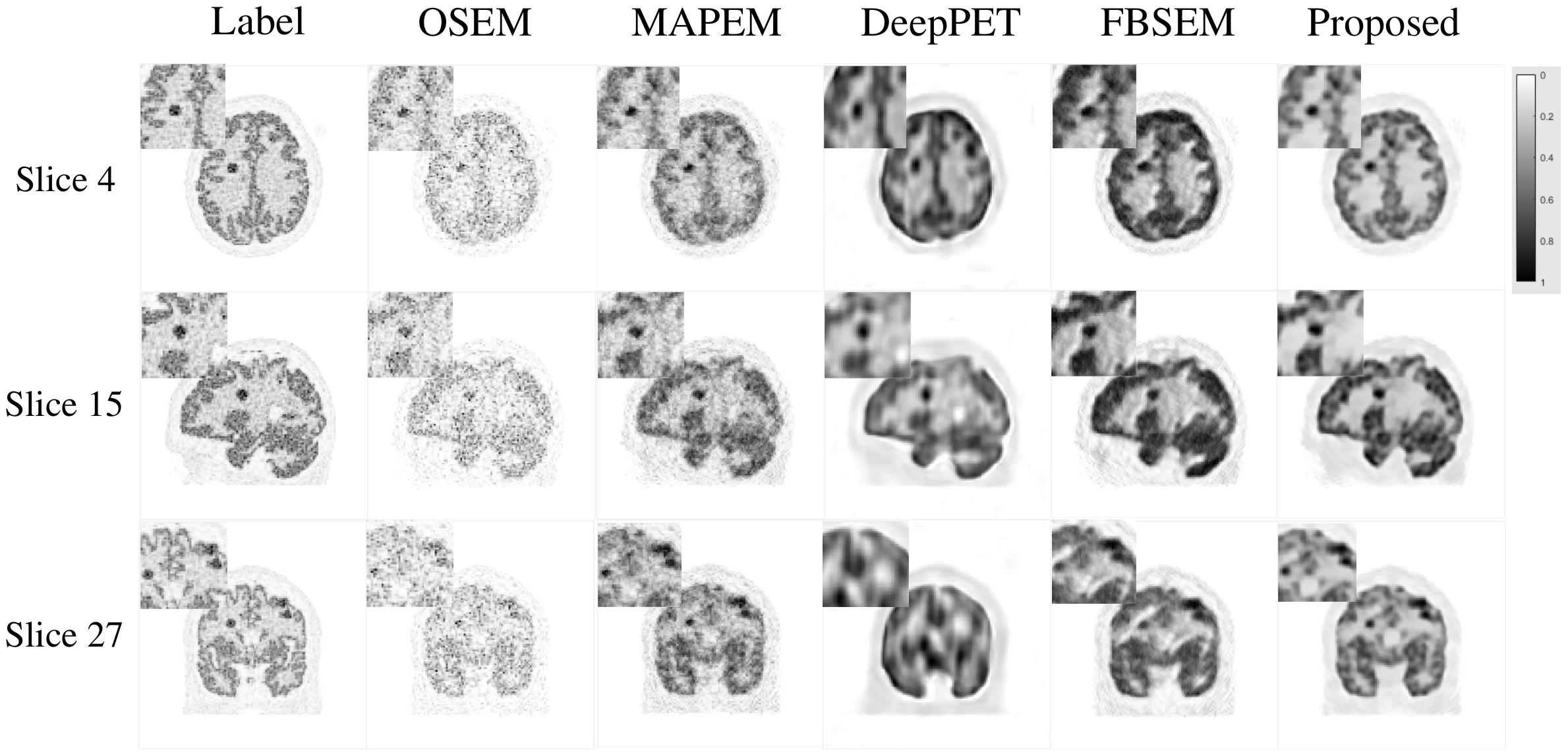}  
\caption{Reconstruction results of OSEM, MAPEM, DeepPET, FBSEM and proposed TransEM on three orthogonal views of one test brain sample.} 
\label{Fig2}
\end{figure}

\section{EXPERIMENT AND RESULTS}
\subsection{Experimental evaluation}
Twenty 3D brain phantoms from BrainWeb~\cite{23} were used to simulate 2D ${}^{18}F$ FDG PET images with the resolution and matrix size of 2.086$\times$2.086$\times$2.031 $mm^3$ and 344$\times$344$\times$127 acquired from a Siemens Biograph mMR. For each phantom, 10 noncontinuous slices were selected from each of the three orthogonal views to generate high count sinograms which were used to reconstruct the label images and low count sinograms with size of 172$\times$252. The system matrix was simulated with Siddon projection~\cite{siddon1985fast}. For high count, $5*10^{6}$ counts and point spread function (PSF) modeling with 2.5$mm$ full width at half maximum (FWHM) Gaussian kernels were used, while $5*10^5$ counts on average and PSF of 4$mm$ were used for low count. The high dose label images were reconstructed from high count sinogram using OSEM algorithm with 10 iterations and 6 subsets. Besides, fifteen hot spheres of radius ranging from 2mm to 8mm were inserted into all phantoms. TransEM has trained with 17 brain samples (510 slices) to map low count sinogram to high dose label PET images, and 2 brain samples (60 slices) for testing and 1 brain sample (30 slices) for validation. To assess reconstruction quality, quantitative comparisons were performed against high dose label images. Both references and reconstructed images were normalized to a maximum of 1. Peak signal to noise ratio (PSNR), structural similarity index (SSIM~\cite{24}) and mean contrast recovery coefficients (MCRC) were calculated.

\begin{equation}
MCRC = \frac{1}{N}\sum\limits_{n = 1}^N {\frac{{\overline {{I_a}} }}{{{I_{true}}}}}
\end{equation}
where $N$ is the number of pictures which contains the simulated tumors, $\overline {{I_a}}$is the average uptake of all the tumor areas in the test phantom.

\begin{figure}[t]
\centering  
\subfigure[PSNR]{
\label{Fig.sub.1}
\includegraphics[width=6cm,height = 4cm]{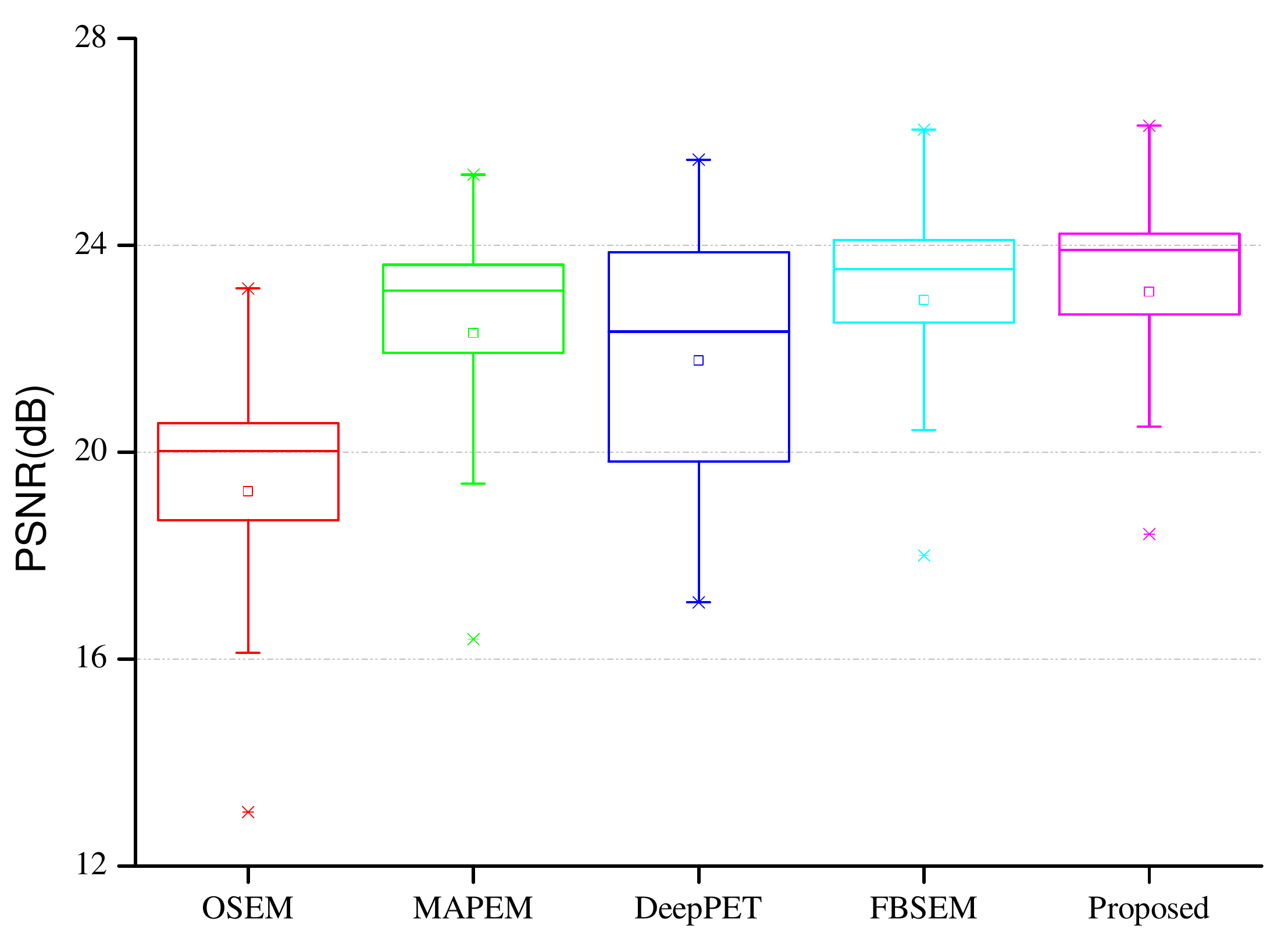}}\subfigure[SSIM]{
\label{Fig.sub.2}
\includegraphics[width=6cm,height = 4cm]{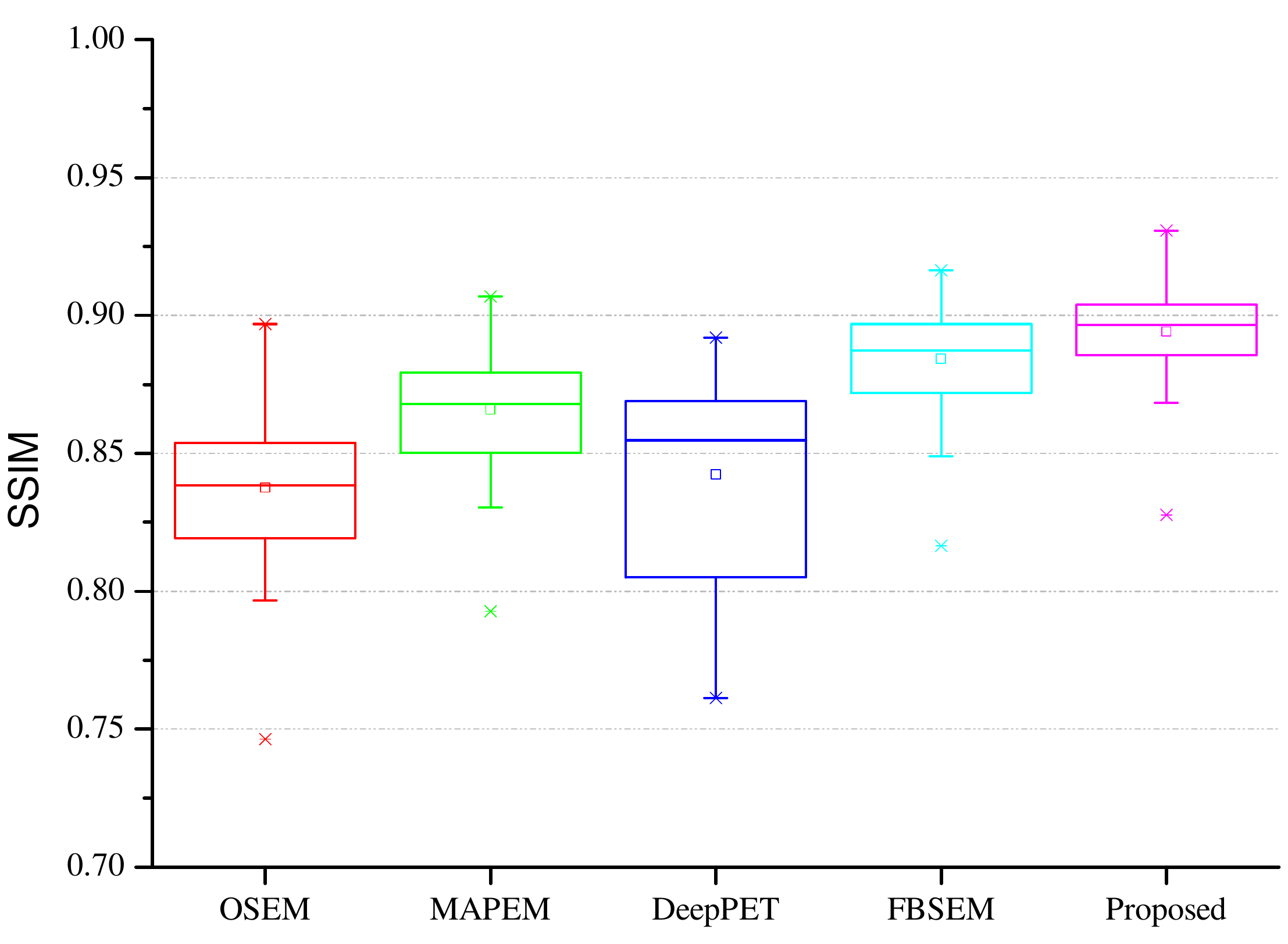}}
\caption{Quantitative image quality(PSNR, SSIM) comparison among different methods .}
\label{Fig3}
\end{figure}

\subsection{Results}
Fig. \ref{Fig2} shows three orthogonal views of the reconstructed brain PET images using different methods. It can be observed that the conventional OSEM algorithm suffers from high level noise. MAPEM reduces noise but always shows over-smooth, resulting in losses of detailed information. As one of the direct learning methods, DeepPET performed not so good. One possible reason is that DeepPET is extremely data-hungry, so poor performance on a small dataset is expected. The FBSEM has a better noise reduction compared to the traditional method OSEM, MAPEM and direct learning method DeepPET, but also has some noises showing up in different regions and some structural information is not well recovered. As seen, the proposed TransEM revealed more cortex structures and preserved edges well compared to other methods. The quantitative results on the test set are demonstrated in Fig. \ref{Fig3} where our proposed method achieves the highest scores among all the methods.

\begin{figure}[t]            
\centering
\includegraphics[width=10cm]{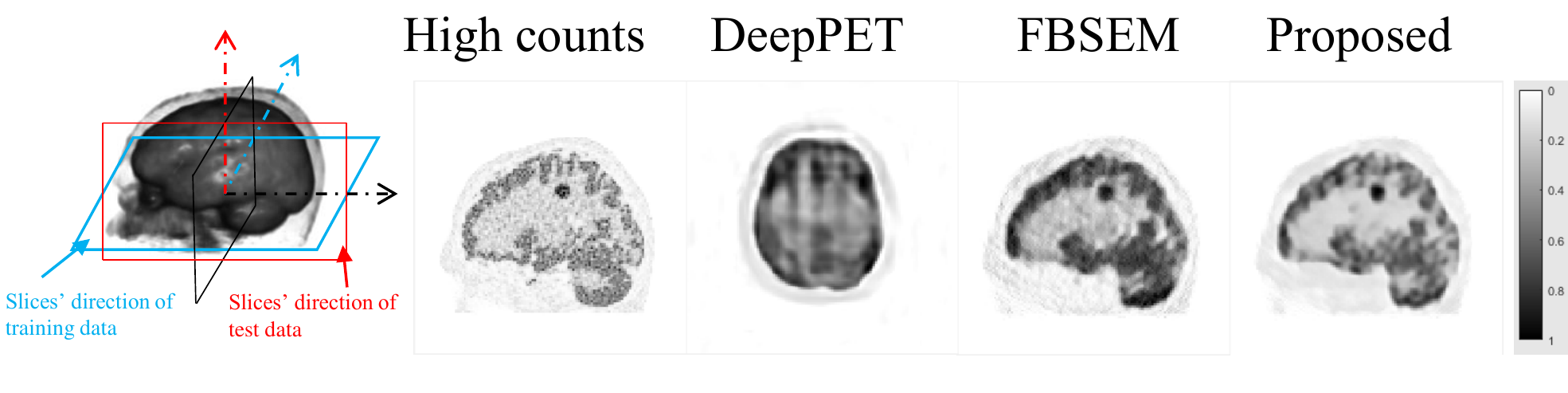}  
\caption{Robustness analysis on the difference slices' direction between the training set and test set. In this experiment, the training slices are selected from the transverse plane, while the test slices are from the sagittal plane.} 
\label{Fig4}
\end{figure}

\begin{table}\scriptsize
    \caption{The PSNR SSIM and MCRC of the test set with different counts level.}
    \begin{tabular}{*{11}{c}}
      \toprule
      \multirow{2}*{Method} & \multicolumn{3}{c}{Counts=1.25e6 (1/4)} & \multicolumn{3}{c}{Counts=5e5 (1/10)} & \multicolumn{3}{c}{Counts=5e4 (1/100)} &\\
      \cmidrule(lr){2-4}\cmidrule(lr){5-7}\cmidrule(lr){8-10}
      & PSNR & SSIM & MCRC & PSNR & SSIM & MCRC & PSNR & SSIM & MCRC\\
      \midrule
      MLEM      & 19.97$\pm$2.69 & 0.86$\pm$0.02 & 0.6852 & 19.24$\pm$2.34 & 0.84$\pm$0.03 & 0.5109 & 15.03$\pm$1.93 & 0.77$\pm$0.03 & 0.1662\\
      MAPEM     & 22.35$\pm$2.26 & 0.88$\pm$0.02 & 0.8187 & 22.30$\pm$2.25 & 0.86$\pm$0.02 & 0.7983 & 17.04$\pm$2.23 & 0.79$\pm$0.03 & 0.3838\\
      DeepPET   & 20.74$\pm$2.05 & 0.82$\pm$0.04 & 0.7005 & 21.77$\pm$2.13 & 0.84$\pm$0.04 & 0.6813 & \pmb{20.69$\pm$2.83} & 0.82$\pm$0.05 & \pmb{0.6690}\\
      FBSEM     & 22.52$\pm$2.00 & 0.88$\pm$0.01 & 0.8448 & 22.94$\pm$1.84 & 0.88$\pm$0.02 & 0.8518 & 19.16$\pm$2.35 & 0.82$\pm$0.03 & 0.5681\\
      Proposed  & \pmb{22.61$\pm$2.00} & \pmb{0.90$\pm$0.01} & \pmb{0.8578} & \pmb{23.10$\pm$1.86} & \pmb{0.89$\pm$0.02} & \pmb{0.8718} & 20.10$\pm$2.47 & \pmb{0.84$\pm$0.03} & 0.5765\\
      \bottomrule
    \end{tabular}
\label{table1}
\end{table}

\subsection{Robustness analysis}
Besides, to analyze the robustness of the proposed TransEM on different low count levels, we have trained DeepPET, FBSEM, and TransEM on 1/4, 1/100 downsampled data. The training label is reconstructed by OSEM with high count(5e6) data. Each experiment involves retraining and testing. As shown in Table \ref{table1}, including 1/10 downsampled data results mentioned above, TransEM beats all comparison methods at different counts except DeepPET in 1/100 downsampled situation, while we would like to emphasize that it looks like DeepPET got pretty good PSNR and MCRC, in ultra-low count situation, due to the lack of physical constraints, the over-fitting of DeepPET is severe and the results are not very reliable which is proved true when we trained the three learning methods with transverse slices and tested with sagittal slices. We selected training slices from the transverse plane and test slices from the sagittal plane to test the generalization ability of three learning-based methods as shown in Fig. \ref{Fig4}. It can be observed that the generalization ability of DeepPET is poor.


\begin{figure}[t]
\centering 
\subfigure[Ablation study on RC]{
\label{Fig.sub.1}
\includegraphics[width=6cm,height = 3.8cm]{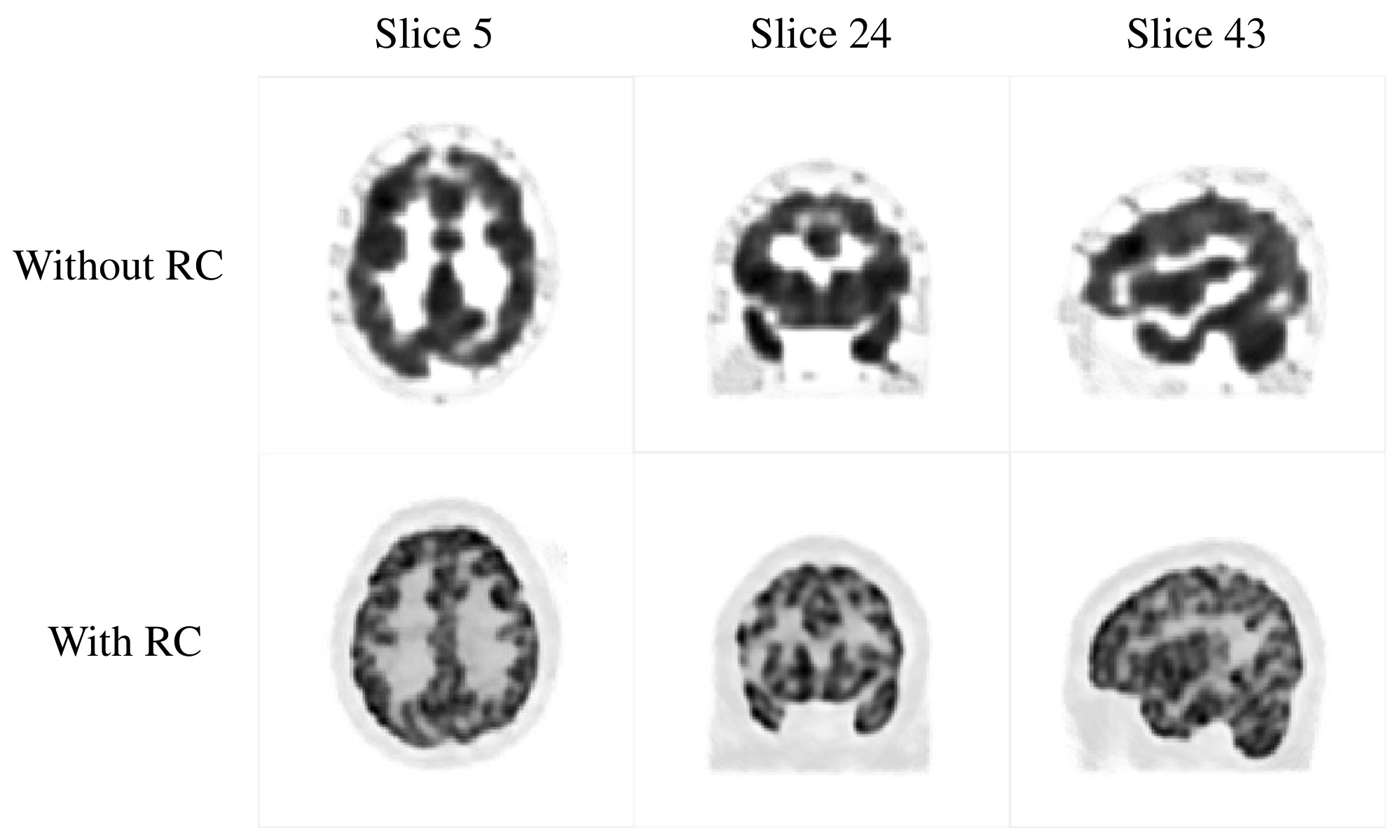}}
\subfigure[Ablation study on Unrolled blocks]{
\label{Fig.sub.2}
\includegraphics[width=5.66cm,height = 3.8cm]{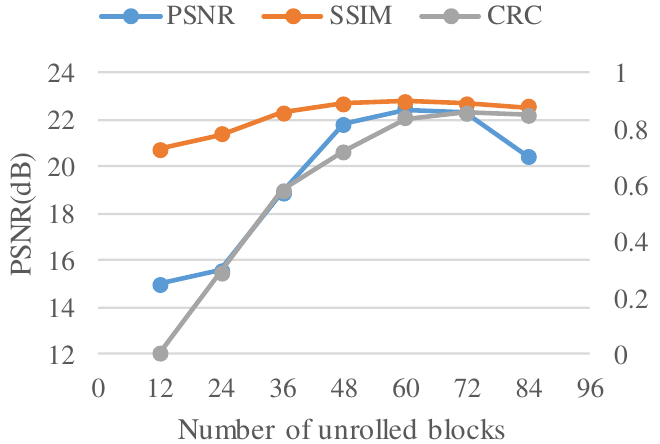}}
\caption{Ablation study on different settings of TransEM.}
\label{Fig5}
\end{figure}

\subsection{Ablation study and discussion}

Figure \ref{Fig.sub.1} shows two residual connection (RC) variants outside STL in RSTR. Without residual connection, the training step is easily falling into sub-optimal solution and is difficult to convergence. The significance of RC also lies in the comparison of reconstruction results.In TransEM proposed in this paper, most of the parameters are learned from training data, however, the number of unrolled blocks is hand-crafted. In this section, the sensitivity of the number of unrolled blocks is analyzed. Due to the limitation of hardware and image size, the number of subsets that we chose is 6, so the number of unrolled blocks is multiples of six. When the number is 60, the TransEM achieves the best performance as shown in Fig \ref{Fig.sub.2}, so the number of unrolled blocks is 60 in the experiment in this paper.

\section{CONCLUSIONS}
In this work, we proposed a model-based deep learning method by unrolling the EM algorithm with residual swin-transformer regularizer for low-dose PET image reconstruction. Simulated human brain data were used in the evaluation. Both quantitative and qualitative results show that the proposed TransEM performs better than the FBSEM, DeepPET as well as traditional OSEM and MAPEM regarding PSNR, SSIM and MCRC. Because lack of clinical PET data currently, future work will focus on more clinical evaluations.

\subsubsection{Acknowledgements}This work was supported in part by the Talent Program of Zhejiang Province (2021R51004) and by the National Natural Science Foundation of China (U1809204).



%
%
%
%

\end{document}